\documentclass[twocolumn,twocolappendix]{aastex63}
\usepackage{multirow}
\usepackage{lipsum}
\usepackage{graphicx,amssymb,amsmath,epstopdf, xfrac}
\usepackage{hyperref}
\usepackage[caption=false]{subfig}

\newcommand{\fuse}{{FUSE}}
\newcommand{\aspera}{{Aspera}}
\newcommand{\luvoir}{{LUVOIR}}
\newcommand{\luvoira}{{LUVOIR-A}}
\newcommand{\luvoirb}{{LUVOIR-B}}
\newcommand{\cetus}{{CETUS}}
\newcommand{\rosat}{{ROSAT}}
\newcommand{\xmmnewton}{{XMM-Newton}}
\newcommand{\chandraobs}{{Chandra X-ray Observatory}}

\newcommand{\OVI}{\ion{O}{6}}
\newcommand{\OVII}{\ion{O}{7}}
\newcommand{\HI}{\ion{H}{1}}

\newcommand{\specialcell}[2][c]{%
  \begin{tabular}[#1]{@{}c@{}}#2\end{tabular}}

\newcommand{\fluxcnt}{counts s$^{-1}$ cm$^{-2}$ sr$^{-1}$}
\newcommand{\fluxerg}{ergs s$^{-1}$ cm$^{-2}$ arcsec$^{-2}$}

\received{}
\revised{}
\accepted{}
\date{January 2021}

\begin{document}
\title{Revisiting FUSE {\OVI} Emission in Galaxy Halos}

\correspondingauthor{Haeun Chung}
\email{haeunchung@arizona.edu}

\author[0000-0002-3043-2555]{Haeun Chung}
\affiliation{University of Arizona, Steward Observatory, 933 N. Cherry Ave., Tucson, AZ 85721, USA}

\author[0000-0001-7936-0831]{Carlos J. Vargas}
\affiliation{University of Arizona, Steward Observatory, 933 N. Cherry Ave., Tucson, AZ 85721, USA}

\author[0000-0002-3131-7372]{Erika Hamden}
\affiliation{University of Arizona, Steward Observatory, 933 N. Cherry Ave., Tucson, AZ 85721, USA}


\begin{abstract}
A significant fraction of baryons in galaxies are in the form of diffuse gas of the circumgalactic medium (CGM). One critical component of the multi-phases of CGM, the so-called ''coronal'' warm-hot phase gas ($\rm 10^{5}-10^{6}$ K) traced by {\OVI} 1031.93, 1037.62 {\AA} resonance lines, has rarely been detected in emission from galaxy halos other than Milky Way. Here we report four additional detections of {\OVI} emission gas in the halos of nearby edge-on galaxies, NGC 4631 and NGC 891, using archival Far Ultraviolet Spectroscopic Explorer data and an updated data pipeline.
We find the most intense {\OVI} emission to be from fields forming a vertical line near the center of NGC 4631, despite the close proximity to the disk of two other fields. 
\added{The detected {\OVI} emission surface brightness are about 1.1$\pm 0.3$ $\times$ $10^{-18}$ to 3.9$\pm 0.8$ $\times$ $10^{-18}$ {\fluxerg}. The spatial distribution of the five 30{\arcsec} $\times$ 30{\arcsec} {\OVI} detection fields in NGC 4631 can be interpreted as}
\deleted{We theorize that the vertically-oriented fields trace a warm-hot phase gas filament that is possibly infalling onto the galaxy. The {\OVI} kinematics of the fields closer to the edges of the disk suggest that those fields sample ejected gas that has shock heated and is cooling before returning to the disk. These results point to}
the existence of filamentary structures of more intense {\OVI} emission superimposed within a \deleted{more} diffuse and faint {\OVI} halo in star-forming galaxies. Volume-filled {\OVI} emission mapping is greatly needed to determine the structure and prevalence of warm-hot gas and the role it plays in the cycling of gas between the galaxy disk and the halo. Finally, we present the sensitivity of future funded and proposed UV missions ({\luvoira}, {\luvoirb}, {\cetus}, and {\aspera}) to the detection of diffuse and faint {\OVI} emission in nearby galaxy halos.

\end{abstract}

\section{Introduction}

In the Lambda Cold Dark Matter ($\Lambda$CDM) paradigm, galaxies form at the centers of dark matter halos out of cooling and condensed gas into a central star-forming disk. While observations of the galaxies at the center of these halos have been conducted for over 100 years, we are only now able to understand the full picture of how gas outside of the central disk plays a key role in the evolution and star formation rate of the galaxy. This full picture of gas is important because a majority ($>80$\%) of the gas in most galaxies will never end up in stars \citep{zaritskyetal17,tumlinsonetal17,Behroozi19}. In these halos, the warm-hot ($\rm 10^{5}-10^{6}$ K) phase of the CGM accounts for more mass than the stars within the parent galaxy. This phase is best traced by the {\OVI} doublet, occurring at $\lambda\lambda$ $1031.93, 1037.62$ \AA~rest frame. This transition is extremely temperature sensitive, and is the strongest line transition that traces warm-hot phase gas. Emission from even this brightest transition occurs at low surface brightness which has challenged observers. 

\begin{figure*}[t]
    \centering
    \subfloat{{\includegraphics[width=0.695\textwidth]{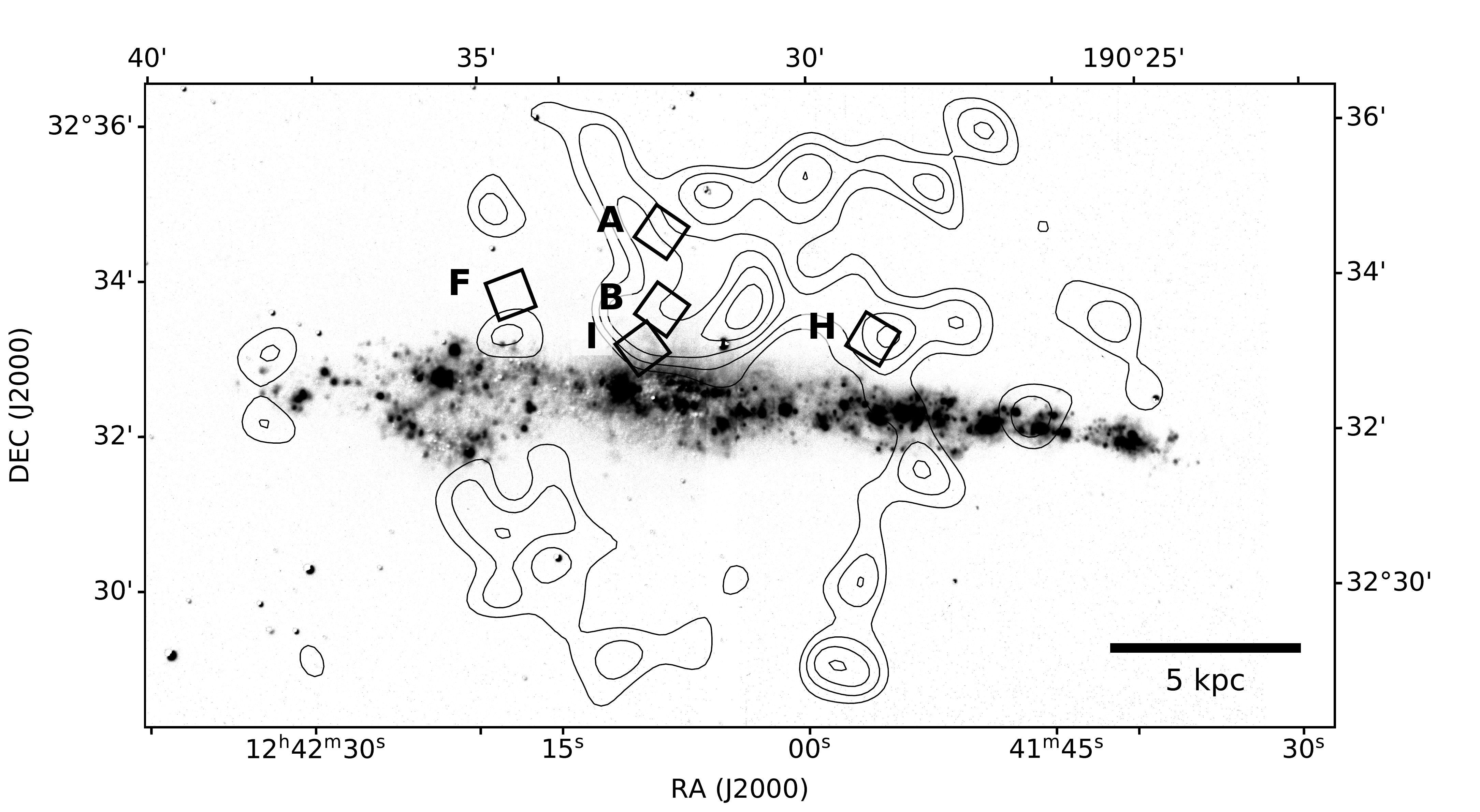} }}
    \subfloat{{\includegraphics[width=0.305\textwidth]{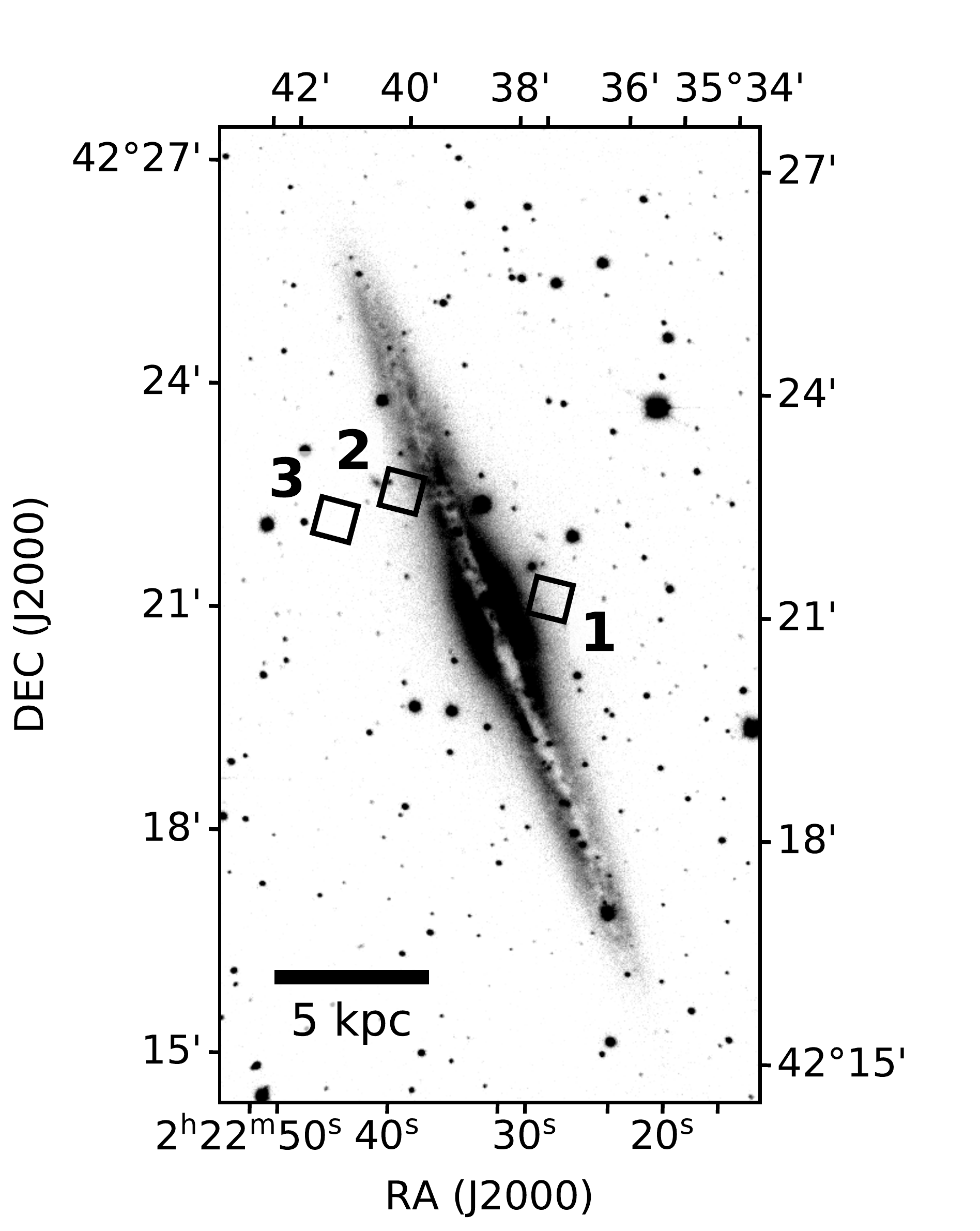} }}
    \caption{
    (Left) H$\rm \alpha$ image of NGC 4631 with soft X-ray band (0.15 - 0.3 keV) contours. Five {\fuse} LWRS observation fields are marked with black squares. NGC 4631-A and NGC 4631-B are previously known {\OVI} detections by \citetalias{2003ApJ...591..821O}. H$\rm \alpha$ image is adopted from SINGS data archive \citep{2003PASP..115..928K}, observed by KPNO 2.1m telescope. Each contour interval equals 1$\sigma$ with the lowest contour at 3 $\sigma$ aboave the local sky background (3.8 $\times$ 10$^{-4}$ counts s$^{-1}$ arcmin$^{-2}$; 1 $\sigma$ = 1.8 $\times$ 10$^{-4}$ counts s$^{-1}$ arcmin$^{-2}$). 7.4 Mpc is used as the distance to NGC 4631 \citep{2011RadburnSmith}.
    (Right) H$\rm \alpha$ image of NGC 891 with three {\fuse} LWRS observation fields, marked with black squares. All three fields were previously known as {\OVI} non-detections (\citetalias{2003ApJ...591..821O}). The image is adopted from \citet{1997UITVi.U..K....C}, observed by Mount Laguana Observatory SDSU 40-inch telescope. 9.1 Mpc is used as the distance to NGC 891 \citep{2011RadburnSmith}.
    }
    \label{fig:ha_images}
\end{figure*}

Due to the historical difficulty of measuring {\OVI} emission, only three studies have presented detections of coronal {\OVI} emission lines outside of the Milky Way: two in galaxy disks and one in the CGM \citep{2003ApJ...591..821O,2007ApJ...668..891G,2016ApJ...828...49H}. \defcitealias{2003ApJ...591..821O}{O03}\citet{2003ApJ...591..821O}(hereafter \citetalias{2003ApJ...591..821O}) provides the most directly relevant results to this work. They searched for {\OVI} emission beyond the disks of two nearby edge-on galaxies---NGC 4631 and NGC 891---using Far Ultraviolet Spectroscopic Explorer ({\fuse}, \citealt{moos2000}) spectra. They detected coronal gas in two fields (30{\arcsec} $\times$ 30{\arcsec} Field-of-view) of NGC 4631 and provided upper limits for NGC 891 (See \autoref{fig:ha_images}).

\begin{deluxetable*}{cccccccc}
\tabletypesize{\footnotesize}
\tablewidth{0pt}
 \tablecaption{{\fuse} Observation Summary\label{tab:obs_summary}}
 \tablehead{ 
  \colhead{\multirow{2}{*}{Field} }
 & \colhead{\multirow{2}{*}{Program ID} }
& \colhead{\specialcell[t]{R.A. \\ (J2000)}}
& \colhead{\specialcell[t]{Decl.\\ (J2000)}}
&\colhead{\specialcell[t]{Exposure Time\\ (Day+Night, sec)}}
&\colhead{\specialcell[t]{Exposure Time\\ (Night, sec)}}
&\colhead{\specialcell[t]{Effective Area\tablenotemark{a}\\ (LiF1-A, cm$^2$)}}
 }
 \startdata
 NGC4631-A &  p1340101 &  12h42m08.8s &  +32d34m36.0s & 21536 & 12737 & 26.9 \\
 NGC4631-B &  p1340201 &  12h42m08.8s &  +32d33m36.0s & 16218 & 11528 & 26.9 \\
 NGC4631-A$\ast$ &  c0570101 &  12h42m08.8s &  +32d34m36.0s &  7153 &  6332 & 24.6 \\
 NGC4631-F &  c0570201 &  12h42m18.0s &  +32d33m48.0s & 16339 & 13573 & 23.2 \\
 NGC4631-F$\ast$ &  c0570202 &  12h42m18.0s &  +32d33m48.0s & 30228 & 18134 & 23.3 \\
 NGC4631-H &  c0570301 &  12h41m56.0s &  +32d33m12.0s & 22866 & 15856 & 23.4 \\
 NGC4631-I &  c0570401 &  12h42m10.0s &  +32d33m06.0s &  9392 &  8447 & 24.3 \\
  NGC891-1 &  b1140101 &   2h22m29.0s &  +42d21m12.0s & 31594 & 27174 & 26.2 \\
  NGC891-2 &  b1140201 &   2h22m40.0s &  +42d22m36.0s & 31440 & 27203 & 26.3 \\
  NGC891-3 &  b1140301 &   2h22m44.8s &  +42d22m12.0s & 18485 & 15666 & 26.2 \\
\enddata
\tablenotetext{a}{Effective area of {\fuse} LiF1-A channel at 1035 {\AA}.}
\end{deluxetable*}

NGC 4631, also known as the Whale Galaxy, is a nearby edge-on barred spiral galaxy with companions NGC 4627 and NGC 4656 \citep{2018Richter}. NGC 4631 has a distance of 7.4 $\pm$ 0.2 Mpc, and, based on {\HI} data, is interacting with several nearby systems \citep{2011RadburnSmith}. NGC 4631 was originally observed with {\fuse} because {\rosat} \added{\citep{1987SPIE..733..519P}} observations indicated a concentration of soft X-ray emission above the plane of the galaxy but correlated with star formation activity in the plane \citep{2002fuse.prop.C057M}. Additional observations of NGC 891 were conducted later, in part because of similar soft X-ray emission detected by {\rosat}. NGC 891 is a nearby, edge-on unbarred spiral galaxy at a distance of 9.1 $\pm$ 0.4 Mpc \citep{2011RadburnSmith}, but unlike NGC 4631, does not have any large companions. NGC 891 is usually compared to our own Milky Way galaxy, in both size and luminosity. There have been numerous studies of the halo of NGC 891 \citep{2005Temple,2018HodgesKluck,2019Qu,2020Das}, which indicate the presence of both neutral hydrogen, a hot halo, and warm ionized gas of less than solar metallicity.  Observations by {\fuse} of NGC 4631 and NGC 891 were conducted between 2000 and 2004.

Here, we present an updated analysis of the {\fuse} data presented in \citetalias{2003ApJ...591..821O}, including several fields which were not included in their original published study. We detect coronal gas via {\OVI} in several sightlines, three of which are aligned in a possible filamentary structure. \deleted{with a decay that can be fitted with an exponential scale height.} In \autoref{sec:data}, we describe the {\fuse} data and our updated analysis. In \autoref{sec:results}, we describe our detections and limits for NGC 4631 and NGC 891. Finally, we discuss the implications of this new analysis in \autoref{sec:discussion}, including the impact on future EUV mission concepts.

\section{Data and Analysis}\label{sec:data}

Our analysis uses existing archival data from the {\fuse} telescope \citep{moos2000}, which was launched in 1999 and operated for 8 years. The data used in this analysis is listed in \autoref{tab:obs_summary}.

\subsection{The {\fuse} Telescope}

{\fuse} was designed to observe in the extreme FUV (905-1187 \AA), and was originally designed to follow up the Copernicus mission \citep{1973Rogerson} with higher sensitivity and spectral resolution to capitalize on the rich variety of astrophysically important lines in the EUV. {\fuse} consists of four co-aligned off-axis parabola telescopes (each 352 $\times$ 387 mm), which each feed light to four Rowland circle type spectrographs. All four channels observe the same field of view, but two channels are optimized for 905-1105 {\AA} \deleted{using SiC coatings on optical surfaces, }while the remaining two are optimized for 1000-1187 {\AA}. \deleted{using Al/LiF coatings on optical surfaces. The simple design and careful use of coatings are both driven by the low reflectance of surfaces in this part of the UV.} The {\fuse} focal plane consisted of two microchannel plate (MCP) detectors \citep{1997Siegmund}. \deleted{which used helical double delay line anodes. Each detector receives light from two channels. The front surface of each MCP has a KBr photocathode to improve performance. The design also reduces gaps in the spectral channels via redundancy.}

\subsection{New unpublished data}

The data presented here consists of 10 {\fuse} low-resolution aperture (LWRS) pointings, which include two galaxies (NGC 4631 and NGC 891) and is a mixture of previously published and unpublished data.
\deleted{The observation of both galaxies was originally motivated by evidence of soft X-ray emission from beyond the galaxy disk observed by {\rosat} \citep{1987SPIE..733..519P}. }The data presented here is publicly accessible.

\subsection{Updated analysis of previously published data}\label{subsec:analysis}
\subsubsection{{\fuse} data}
\begin{figure*}
\centering
\includegraphics[width=1.0\textwidth]{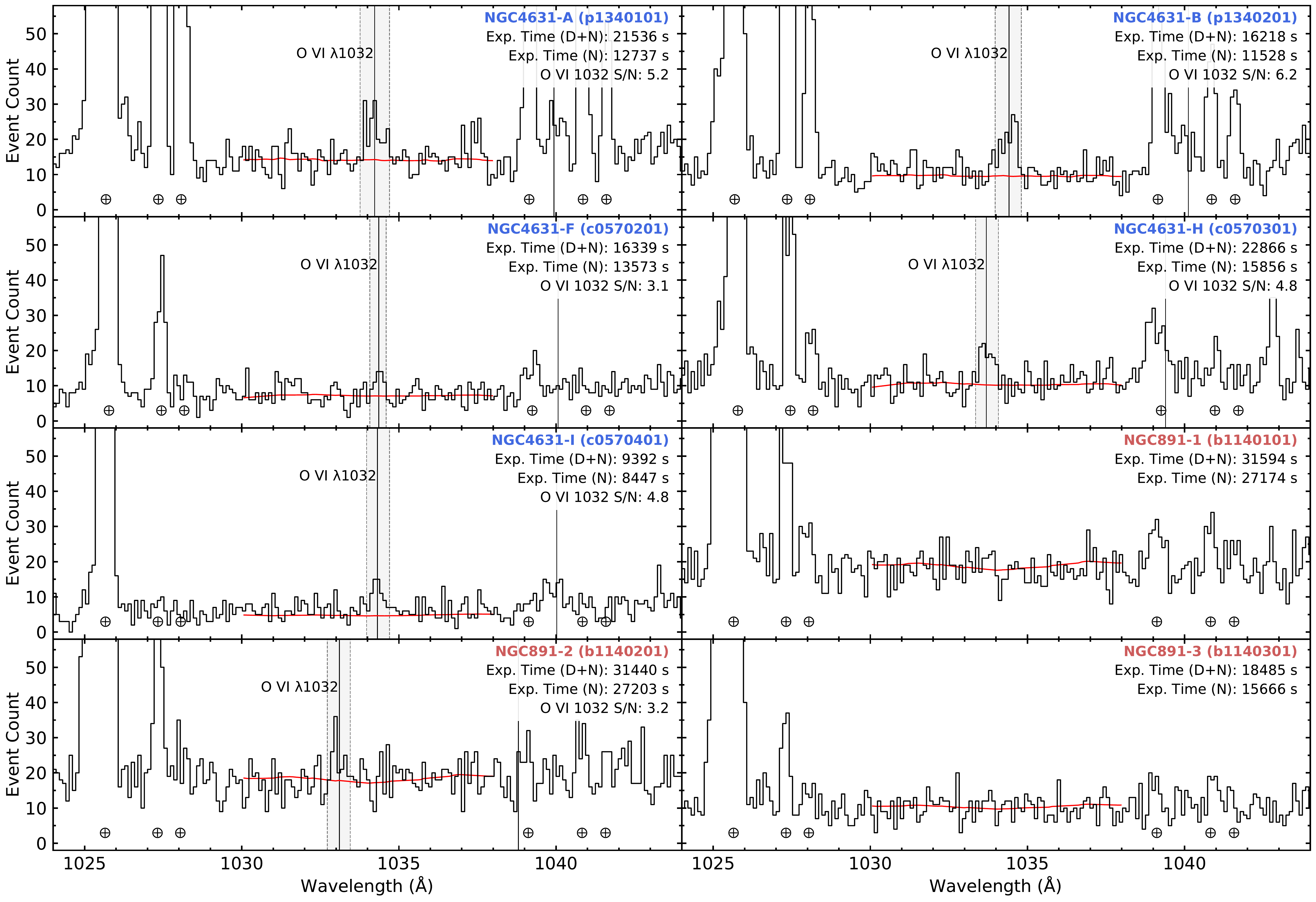}
\caption{{\fuse} LWRS Day+Night spectra of NGC 4631-A, B, F, H, I fields and NGC 891-1, 2, 3 fields. The spectra were binned by 16 pixels. {\OVI} 1032 {\AA} line centers derived from kinematic fitting are plotted as vertical black line, as well as the corresponding {\OVI} 1038 {\AA} location. Background count spectrum around {\OVI} 1032 {\AA} line is plotted in red. The width of the emission line extraction window is shown with the grey shaded area. Airglow line locations are marked with the Earth symbol. Exposure time (Day+Night and night only) and SNR of the {\OVI} line emission are marked at each panel. \added{Note that the background count spectrum is proportional to the exposure time.}} \label{fig:spec_all}
\end{figure*}

The {\fuse} raw data is downloaded from the Canadian Astronomy Data Centre. The data is further processed by the latest release of \texttt{CalFUSE} (\texttt{v3.2.3}, \citealt{2007PASP..119..527D}). 
We use the default pulse height range (between 2 and 25\added{, inclusive}) of \texttt{CalFUSE v3.2.3}, which is wider than the range used by \citetalias{2003ApJ...591..821O} (between 4 and 15\added{, inclusive}). \added{Using the default pulse height range is strongly advised by the pipeline developers, since the use of narrower pulse height range can lead to a significant flux loss \citep{2007PASP..119..527D}.} The extracted spectrum from exposures of each observing program ID are combined using {\fuse} Tools in C\footnote{https://archive.stsci.edu/fuse/analysis/toolbox.html} (\texttt{idf\_combine}, \texttt{cf\_bad\_pixels}, \texttt{bpm\_combine}, and \texttt{cf\_extract\_spectra}), following the recommendation from the web instruction. This is to best estimate the background level when the targets are faint. 
\replaced{we use wavelength (WAVE) and the dead-time corrected counts values (WEIGHTS) in the combined spectrum.}{Among the output of the pipeline, we use wavelength (WAVE) and the dead-time corrected counts values (WEIGHTS) as a spectrum (\autoref{fig:spec_all}). The background count spectrum (BKGD) is used to calculate the signal (Signal = WEIGHTS - BKGD). The raw counts spectrum (COUNTS) is also used for generating Poisson random noise added spectrum which is described in the following paragraph.}

Before measuring the SNR of the {\OVI} 1032 {\AA} emission, we fitted the {\OVI} 1032 {\AA} line emission by a \added{line profile model to measure its kinematics and determine the optimal extraction window size. The model profile is a} convolution of three functions 1) a Gaussian function as a {\OVI} emission line model, 
2) a function representing the {\fuse} point-spread-function (PSF), assumed to be Gaussian with a Full-Width Half-Maximum (FWHM)=0.021 {\AA} (corresponds to the spectral resolution at the central wavelength of LiF1-A channel (1034.7 {\AA}, R$\sim$21000)), and 3) a top-hat function corresponding to the width of the LWRS slit (30{\arcsec}). Four parameters (Amplitude and the FWHM of the {\OVI} emission model, a center of the convolved function, and a constant) are set as free parameters. \replaced{In order to estimate the uncertainties from this low signal to noise ratio spectra,
 we have generated 1000 Poisson random-noise added spectra for each spectrum. We fit the convolved function to each noise added spectrum using a non-linear least square fitting method (\texttt{scipy.optimize.curve\_fit}).}{
 For each spectrum, we have generated 1000 Poisson random-noise added spectra (according to the SNR at each spectrum bin). The convolved function is fitted to each noise added spectrum using a non-linear least square fitting method (\texttt{scipy.optimize.curve\_fit}).} 
  An average and standard deviation of the fitted parameters (line center and FWHM) from 1000 fittings are determined as the parameter value and error. \added{We put an example of this line fitting to NGC 4631-A spectrum in Appendix A.} The fitting was not attempted for the spectra with no visible {\OVI} 1032 {\AA} emission line (NGC 891-1 and 3).

After the fitting, we measured SNR of the {\OVI} 1032 {\AA} emission line. The size of the extraction window (in the wavelength dimension) is determined as the width corresponding to 95\% of the area of
the final convolved function. The final convolved function is the convolution of the three functions (two Gaussians and one top-hat), adopting the {\OVI} emission FWHM value from the fitting result.
The width of the extraction windows are shown in the grey shaded region in \autoref{fig:spec_all}. The {\OVI} signal count is calculated by the difference between the total event counts and the total background counts within the extraction window. The {\OVI} signal count is converted to flux per unit angular area in both {\fluxcnt} (Line Unit, LU) and {\fluxerg}. The {\fuse} effective area at the time of observation is estimated by the interpolation between the two closest {\fuse} effective area data to the epoch time of each observation. The time-dependent spectral effective area curves are available from the \texttt{CalFUSE} calibration files. The mean effective area values at 1035 {\AA} are presented at \autoref{tab:obs_summary}.

\subsubsection{{\rosat} data}
\added{The NGC 4631 {\fuse} field points were chosen based on this {\rosat} observation. Note that higher-resolution X-ray data is available from more recent missions such as {\xmmnewton} or {\chandraobs}. Here we use only {\rosat} data to show the correlation between the soft X-ray band contour (0.15-0.3 keV) and the {\fuse} field points.} We re-analyzed the X-ray observation made by the {\rosat} Position Sensitive Proportional Counter (PSPC, \citet{1982AdSpR...2d.241T}) on NGC 4631. The result was previously published by \citet{1995ApJ...439..176W}. 

Here we reproduce that result and show the soft X-ray band contour in \autoref{fig:ha_images}. The {\rosat} data of NGC 4631 (sequence id: rp600129a00, rp600129a01) is downloaded from the {\rosat} Data Archive at Goddard Space Flight Center. First, we examined the event rate extension (EVRATE) from the ancillary file (\texttt{*\_anc.fits}) to manually define the good time intervals. This step is necessary to exclude bad time intervals contaminated by scattered solar X-rays. The remaining good time intervals are 14.2 ks from rp600129a00 and 3.5 ks from rp00129a01. With the remaining data (total 17.8 ks), we construct a 3D X-Ray data cube (X, Y, and pulse height) using the events list (STDEVT) from the basic file (\texttt{*\_bas.fits}). The final soft-band X-ray image is made from the pulse height-invariant channels interval between 20 to 41, as described in \citet{1995ApJ...439..176W}. This interval corresponds to the energy range 0.15-0.3 keV. The contour in \autoref{fig:ha_images} is constructed with a pixel size of 5{\arcsec} and smoothed by a FWHM=40{\arcsec} Gaussian.

\begin{deluxetable*}{cccccccccc}
\tabletypesize{\footnotesize}
\tablewidth{0pt}
 \tablecaption{O VI Derived Parameters\label{tab:line_params}}
 \tablehead{ 
 \colhead{\multirow{2}{*}{Field} }
 & \colhead{\multirow{2}{*}{Program ID} } 
& \colhead{\specialcell[t]{z \\ (kpc)}}
&\colhead{\specialcell[t]{$\rm I_{1032}$\\ (cnts/s/cm$^2$/sr)}}
&\colhead{\specialcell[t]{$\rm I_{1032}$ ($\times$ 10$^{-18}$\\ ergs/s/cm$^2$/arcsec$^{2}$)}}
&\colhead{\specialcell[t]{Signal, Background \\ (counts, counts)}}
&\colhead{\specialcell[t]{Line Center \\ ($\rm\AA$)}}
&\colhead{\specialcell[t]{Line FWHM \\ ($\rm\AA$)}}
 }
 \startdata
 NGC4631-A &  p1340101 & 4.9 &  6000 $\pm$ 1200 &  2.7 $\pm$ 0.5 &  73, 127 &  1034.23 $\pm$ 0.09 &  0.56 $\pm$ 0.25 \\
 NGC4631-B &  p1340201 & 2.7 &  8300 $\pm$ 1300 &  3.7 $\pm$ 0.6 &   76, 76 &  1034.41 $\pm$ 0.07 &  0.41 $\pm$ 0.19 \\
 NGC4631-F &  c0570201 & 2.8 &  2900 $\pm$ 1000 &  1.3 $\pm$ 0.4 &   23, 35 &  1034.36 $\pm$ 0.08 &  0.15 $\pm$ 0.15 \\
 NGC4631-H &  c0570301 & 2.4 &  4700 $\pm$ 1000 &  2.1 $\pm$ 0.5 &   53, 71 &  1033.69 $\pm$ 0.06 &  0.34 $\pm$ 0.12 \\
 NGC4631-I &  c0570401 & 1.6 &  8600 $\pm$ 1800 &  3.9 $\pm$ 0.8 &   41, 32 &  1034.32 $\pm$ 0.07 &  0.39 $\pm$ 0.29 \\
  NGC891-1 &  b1140101 & 2.0 &         $<$ 2200\tablenotemark{a} &        $<$ 1.0\tablenotemark{a} &  & & \\
  NGC891-2 &  b1140201 & 1.5 &   2300 $\pm$ 700 &  1.1 $\pm$ 0.3 &  40, 124 &  1033.11 $\pm$ 0.17 &  0.36 $\pm$ 0.49 \\
  NGC891-3 &  b1140301 & 4.1 &         $<$ 2700\tablenotemark{a} &        $<$ 1.2\tablenotemark{a} &  & & \\
\enddata
\tablenotetext{a}{These values are 3$\rm \sigma$ upper limit from 1 {\AA}-wide extraction window at around 1033.3 {\AA}.}
\end{deluxetable*}

\section{Result}\label{sec:results}
\subsection{NGC 4631}
\autoref{fig:spec_all} shows the resulting spectra (Day+Night) at the wavelength around {\OVI} emission lines. The known airglow lines are visible in most of the spectra ({\HI} 1025.7, OI 1025.8, 1027.4, 1028.2, 1039.2, 1040.9, 1041.7 {\AA}; Earth symbols in \autoref{fig:spec_all}). The {\OVI} 1038 {\AA} line is not identifiable in most cases because it is blended with the airglow lines. However, the {\OVI} 1032 {\AA} line is visible at the airglow-free window between 1030 to 1038 {\AA}. Among the {\fuse} observations in \autoref{tab:obs_summary}, NGC 4631-A$\ast$ (Program ID: c0570101) is not presented because of the short exposure time, although the data shows an {\OVI} 1032 {\AA} emission signal with SNR=2.7. Another observation, NGC 4631-F$\ast$ (Program ID: c0570202) is highly contaminated by unknown scattered light so the data is discarded from further analysis. 

From the spectra, we derived {\OVI} 1032 {\AA} intensity, signal counts, background counts, line center and line FWHM values. Those values are presented in \autoref{tab:line_params}. In this work, {\OVI} 1032 {\AA} emission signal is detected from all five NGC 4631 fields, two (A, B) with SNR greater than 5, two (H, I) with SNR = 4.8, and the other (F) with SNR = 3.1. \added{We would like to note that the line detection at NGC 4631 field F is only a marginal detection.} The derived {\OVI} 1032 {\AA} intensity value at field A (6000 $\pm$ 1200 {\fluxcnt}) and field B (8300 $\pm$ 1300 {\fluxcnt}) are higher than the ones reported by \citetalias{2003ApJ...591..821O} (4600 $\pm$ 1000 {\fluxcnt} for field A and 8000 $\pm$ 1000 {\fluxcnt} for field B). 

In particular, the signal from field A is estimated to be 30 \% higher than the result from \citetalias{2003ApJ...591..821O}. This is because of the differences between the data reduction pipeline used in this work (\texttt{CalFUSE v3.2.3}) and \citetalias{2003ApJ...591..821O} (\texttt{CalFUSE v2.0.5}), such as pulse height range, extraction window height, good time interval definition, wavelength calibration, etc. Among those differences, the difference in the pulse height range \replaced{would be}{is} the main cause of the difference in {\OVI} intensity. \citetalias{2003ApJ...591..821O} used a narrow pulse height range to reduce the background noise level. However, years later, it was reported that the use of a narrow pulse height range could result in flux losses \citep{2007PASP..119..527D}.

\added{To verify whether the pulse height range is the major contributor of the difference, we have processed the raw data of NGC 4631 field A and B with the narrower pulse height range (4 and 15, inclusive, same as \citetalias{2003ApJ...591..821O}) using the latest pipeline (\texttt{CalFUSE v3.2.3}). The {\OVI} 1032 {\AA} intensity at the two fields are measured as 4800 $\pm$ 1000 {\fluxcnt} for field A and 8200 $\pm$ 1300 {\fluxcnt} for field B, which are comparable to the previous estimation of \citetalias{2003ApJ...591..821O}.
}

The line center and line FWHM values of field A and field B agree with the result of \citetalias{2003ApJ...591..821O}. Note that the derived FWHM value assumed a fully illuminated LWRS (30{\arcsec} $\times$ 30{\arcsec}). Therefore the derived FWHM values can be considered as a lower limit.

\subsection{NGC 891}
\autoref{fig:spec_all} shows the Day+Night spectra of three NGC 891 fields (1,2,3) observed by {\fuse} LWRS. Previously all three fields were considered {\OVI} 1032 {\AA} non-detections. In this re-visit of the {\fuse} spectra, we find a 3.2$\rm \sigma$ signal of {\OVI} 1032 {\AA} line emission at field 2, \added{although this is only a marginal detection.} Field 1 and 3 still show non-detections but the lowest 3$\rm \sigma$ upper limit ($<$ 2200 \fluxcnt, field 1) is slightly increased compared to \citetalias{2003ApJ...591..821O} ($<$ 2000 \fluxcnt, the same field 1).

The derived {\OVI} 1032 {\AA} emission intensity, signal counts, background counts, line center and line FWHM values at field 2 are reported in \autoref{tab:line_params}, along with the 3$\rm \sigma$ upper limit intensity at field 1 and 3. The line of sight velocity and line FWHM at field 2 are 342 $\pm$ 51 km/s and 106 $\pm$ 142 km/s, respectively.

\section{Discussion and Conclusions}\label{sec:discussion}

Here we discuss the potential implications of these {\OVI} detections on our understanding of the distribution and kinematics of warm-hot coronal gas. We also discuss how this crucial component of galaxy halos can be best observed with future funded or proposed mission concepts.

\subsection{Implications for O VI distribution}

\deleted{To date,}The expected morphological distribution of {\OVI} emitting warm-hot gas is debated. Simulations from \citet{corliesetal16} show a \deleted{highly localized} filamentary and clumpy structure in the simulated {\OVI} distribution that largely persists from $z=1$ to $z=0$ (e.g., Figure 6, therein). The empirical work done by the COS-Halos survey in absorption line systems shows large covering fractions for {\OVI} absorption within $75$ kpc of star-forming galaxies \citep{Werketal13}. This implies a more uniformly diffuse {\OVI} morphology in actively star-forming galaxies, \added{not the filaments show in simulations}.

\begin{figure}
\includegraphics[width=\linewidth]{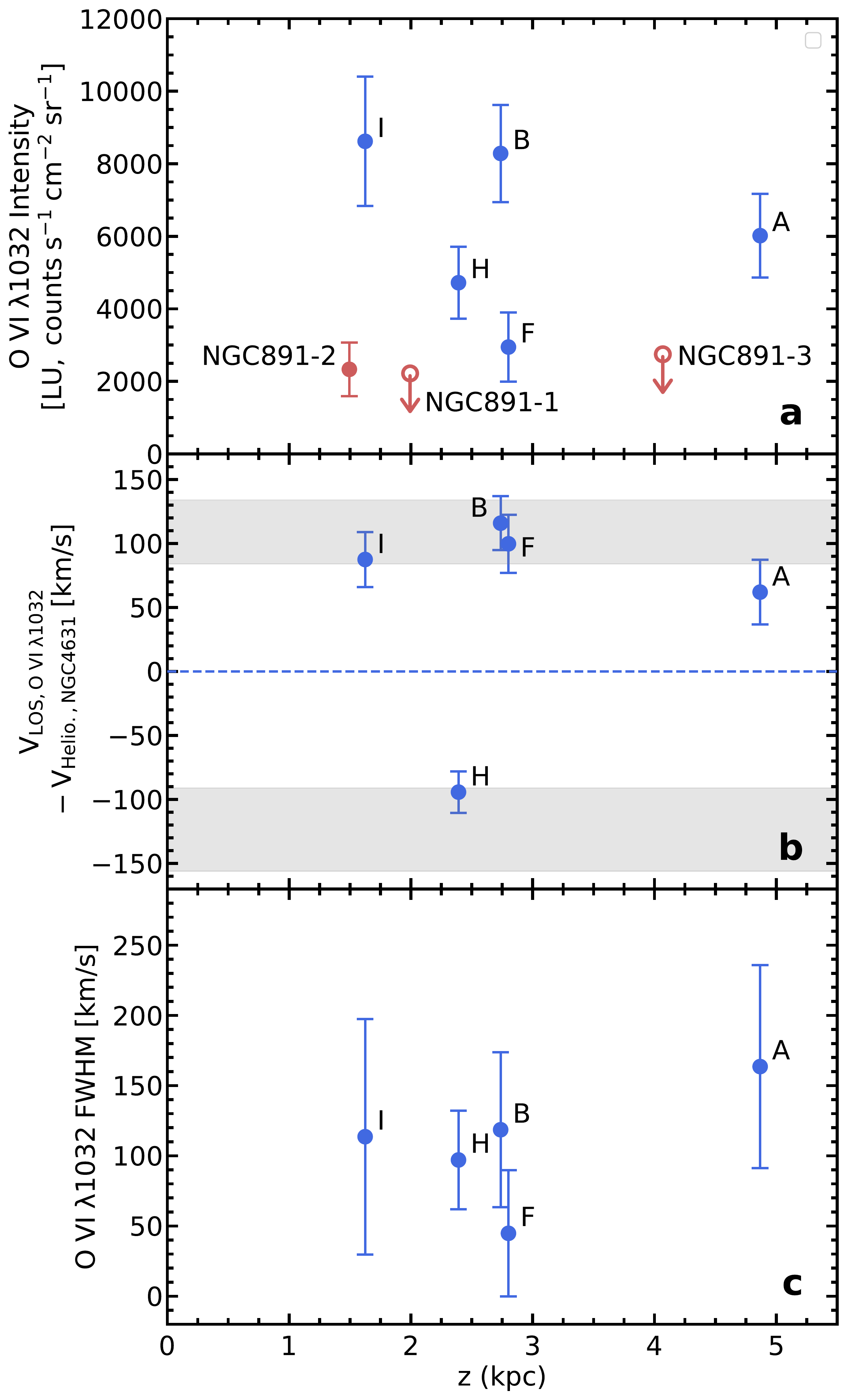}
\caption{(a) {\OVI} 1032 {\AA} line emission intensity, (b) line-of-sight velocity of {\OVI} 1032 {\AA} with respect to the galaxy center (606 km/s; Heliocentric velocity to NGC 4631, \citet{1991rc3..book.....D}), and (c) FWHM of {\OVI} 1032 {\AA}, all as a function of the vertical height from the disk. \replaced{All points are from the NGC 4631 fields, except for the red point in (a) which is from NGC 891 field 2.}{All blue points are from the NGC 4631 fields. One red filled point is from NGC 891 field 2 and the other two red points indicate 3$\sigma$ upper limit at NGC 891 field 1 and 3.} \deleted{The light-blue curve in (a) is an exponential function fitted by using only three points (NGC 4631 field A, B, and I), which are aligned along a line perpendicular to the galaxy disk.} \added{Horizontal gray bars are shown in panel (b) that represent the range in {\HI} velocity within the disk below the FUSE pointings in NGC 4631, as measured using HALOGAS data in \citet{wuthesis}.}
} \label{fig:ovi_values}
\end{figure}

The vertical variation of intensity, velocity, and {\OVI} FWHM for our sample fields are shown in \autoref{fig:ovi_values}. The results of this study provide preliminary evidence for a filamentary or clumpy {\OVI} morphological distribution in the inner CGM. NGC 4631 field\added{s} F \added{and H} show\deleted{s} the weakest {\OVI} emission, despite \replaced{its}{their} proximity to the disk and an {\HI} supershell \citep{rand93}. This stands in contrast to the stronger detections along a nearly vertical line connecting NGC 4631 fields A, B, and I, \added{which we assert provides} \deleted{providing} evidence for a \deleted{discrete} {\OVI}-emitting filament. \deleted{A large-scale filamentary structure in {\OVI} is possibly tied to cooling inflowing matter originating from the intergalactic medium, rather than from outflowing gas from superbubbles or AGN feedback within the disk.}

\deleted{It is possible that filamentary structures exist superimposed} \added{Filamentary structures may be superimposed} within a diffuse {\OVI} halo that is less-intensely emitting, and below the detection threshold for many of the fields presented in this study. Filaments would then be the interface of infalling flows with a diffuse halo of \textit{fountain} gas primarily originating from the disk. This interpretation brings the results from this study, \citet{corliesetal16}, and \citet{Werketal13} into agreement. \deleted{Indeed, }The simulated {\OVI} halo in Figure 6 from \citet{corliesetal16} would suggest this is the most likely scenario. Volume-filled mapping observations of {\OVI} in emission are \deleted{clearly} needed for a more definitive view on the morphology of warm-hot halos in nearby galaxies.

\deleted{Interestingly, }The \deleted{hypothesized} filament structure along fields A, B, and I lies above a well-documented molecular gas outflow near the disk \citep{irwinetal11,rand00}. \deleted{Also, regions F and H both lay above {\HI} supershells found by \citet{rand93}. Though all of these cases represent matter being vertically ejected, }The strength of the {\OVI} detections are stronger along the central filament than near the periphery of the disk. This is potentially due to the outflowing gas near the central molecular outflow interacting with infalling material within the filament, heating the gas, and enhancing {\OVI} emission. Alternatively, the giant magnetic rope features discovered in \citet{moraetal19} may be altering the distribution of, or otherwise interacting with, the ionized {\OVI}-emitting gas.

\deleted{An exponential scale height model is fit to the vertical intensity distribution of {\OVI} emission using fields A, B, and I, and shown as a blue line in panel (a) of \autoref{fig:ovi_values}. The estimated {\OVI} scale height is $16.9\pm4.4$ kpc. This scale height is roughly an order of magnitude larger than previously-measured ionized gas scale heights from the EDGE-CALIFA Survey \citep{levyetal19}, implying a much more vertically-extended {\OVI}-emitting warm-hot halo.}

\added{We find stronger {\OVI} emission in NGC 4631 than in NGC 891. It is possible that this discrepancy is due to the difference in the virial temperature of the two galaxies. The mass of NGC 4631 is closer to that of an L* galaxy, and thus its virial temperature would be closer to the temperatures required for {\OVI} emission. Higher mass systems, like NGC 891, may have a larger fraction of their CGM in the higher ionization `hot' phase, probed by {\OVII} and other X-ray metal lines. }

\subsection{Implications for O VI kinematics}

Panel (b) of \autoref{fig:ovi_values} shows the vertical kinematic distribution of {\OVI} emitting gas in each of the observed fields in NGC 4631, relative to the galaxy's systemic velocity. NGC 4631 field H is the only field on the rotationally approaching side of the galaxy, while the other fields are all on the receding side. \added{The rotational direction of the underlying disk of NGC 4631 was checked using public {\HI} data from the HALOGAS survey \citep{healdetal11}. The {\HI} data verifies that the disk rotation matches the roation direction implied by each {\OVI} detection above the disk in NGC 4631.  }

Though there are only five fields included, the vertical {\OVI} kinematics can be compared to other inner-halo phase kinematics. Gas kinematics above spiral galaxy disks are expected to kinematically `lag' the underlying disk rotation due to angular momentum conservation of ejected disk matter \citep{bregman80}. \citet{wuthesis} obtained vertical slit spectroscopy using the Apache Point Observatory (APO) 3.5-m Telescope to study the vertical kinematics of H$\alpha$ emission in NGC 4631. The {\OVI} filament along fields A, B, and I most closely correspond to slit 2 in the central field from \citet{wuthesis}. The H$\alpha$ kinematics along that slit show roughly the same velocity magnitude as we find along fields A, B, and I (within uncertainties). However, for recently outflowing gas, one would expect the largest rotational velocity to be in the field closest to the disk -- it will lag the disk rotation more intensely the further out it travels. It is not clear that that behavior is found along fields A, B, and I. \deleted{, and hence we tentatively favor the infalling hypothesis for the filament.}  

Fields F and H, which lay further \deleted{out} from the galaxy's center radially, both show velocities very close to the rotation curve at their radial distance, as inferred by an {\HI} position-velocity diagram along the major axis (data from \citealt{healdetal11}; presented in \citealt{wuthesis}). This regular rotation further supports the hypothesis that the {\OVI} emitting gas from fields F and H represent originally shock-heated ejected matter that is cooling and falling down back onto the galaxy's disk. \added{Since the velocity at points F and H do not show a lag to the underlying disk rotation (within uncertainties), this gas may be outflowing and recently ejected from the disk.} We note that these inferences are made using a small number of available data points, and that a spectroscopic, volume-filled {\OVI} emission mapping campaign is needed for more robust conclusions.  

Panel (c) of \autoref{fig:ovi_values} includes the measured {\OVI} FWHM values as a function of vertical distance. The FWHM values are relatively uniform and show no clear trend or behavior. 

\begin{figure}
\includegraphics[width=\linewidth]{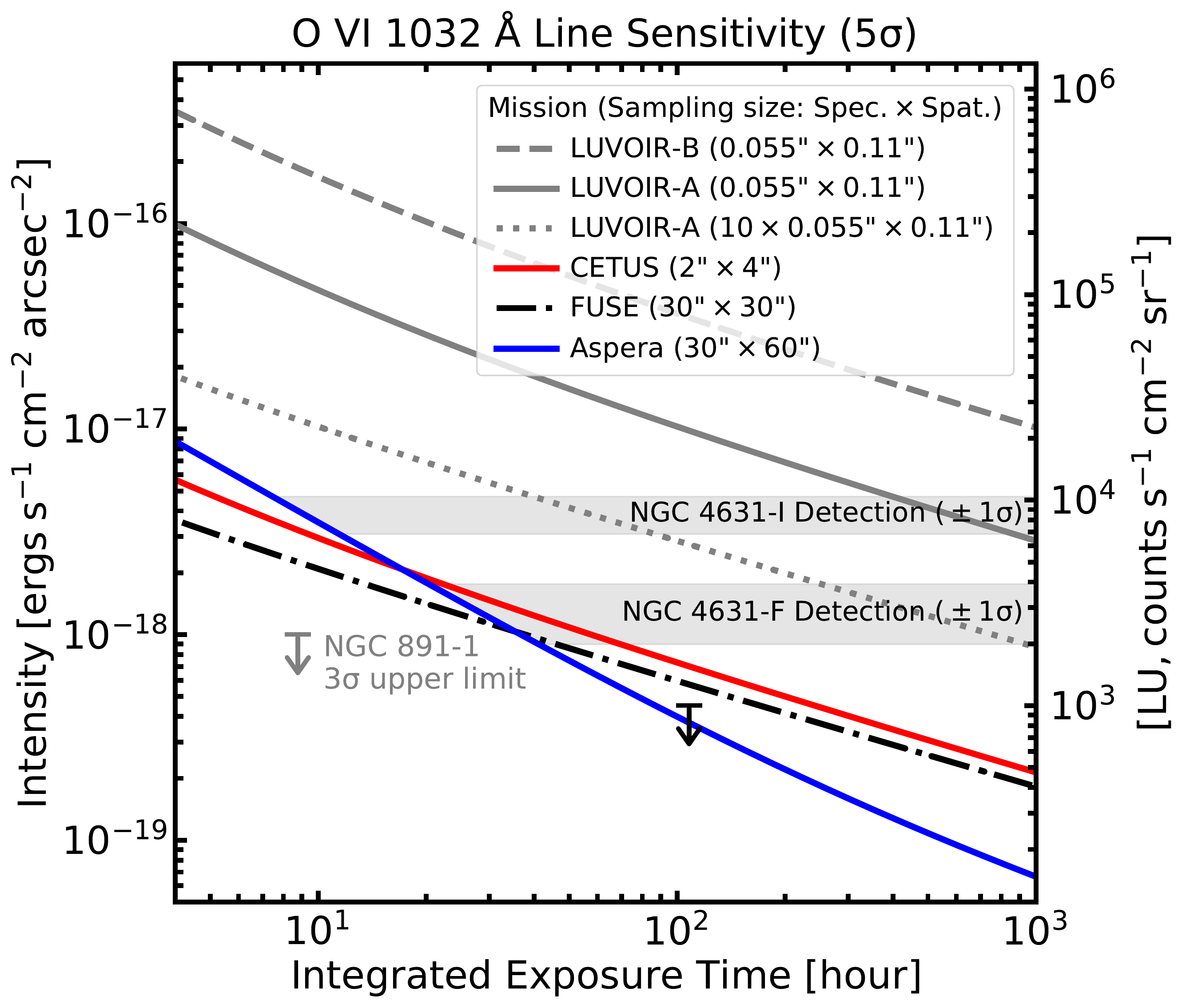}
\caption{
{\OVI} 1032 {\AA} line emission sensitivity (5$\rm \sigma$, per sampling element) of the past ({\fuse}) and future ({\luvoira}, {\luvoirb}, {\cetus}, and {\aspera}) UV missions. Each line is expected sensitivity vs. exposure time of each mission for a given sampling size with a specific configuration (\autoref{tab:snrbreakdown}, \autoref{app:sensitivity_calc_description}). 
Two of {\OVI} emission detection ranges ($\rm \pm1\sigma$) and one non-detection upper limit ($\rm 3\sigma$) are also shown.
The dotted grey line represents the sensitivity of {\luvoira} with co-added {\OVI} emission from ten 0.054{\arcsec} $\times$ 0.108{\arcsec} micro shutter apertures. The black arrow indicates the sensitivity requirement of {\aspera} mission at the exposure time of 4.5 days. 
} \label{fig:sensfuture}
\end{figure}

\subsection{Implications for {\OVI}  emission detection/mapping with future missions}

Studies of galaxy halos using \textit{HST-COS} have yielded a significant amount of information and insight into the composition of various CGM components \citep{Werketal13}, including the seemingly large covering fraction of {\OVI}. But the limitations of relying on infrequent background QSOs to probe foreground galaxy halos is clear with this work. The limited spatial coverage by {\fuse} of NGC 4631 and NGC 891 demonstrate that full coverage of a galaxy halo is needed to definitively conclude if there is a uniform or filamentary distribution of {\OVI} gas. Here we briefly discuss the possibilities of doing wide-field, imaging spectroscopy over a full galaxy halo, comparing several proposed and funded mission concepts.

The signature of {\OVI} is extremely faint and spread out over a large angular area for nearby galaxies (10s of arcmin), which motivates a wide field but high throughput telescope with moderate spectral resolution. Moderate spectral resolution will allow us to distinguish between nearby geocoronal lines and identify the velocity shift of the {\OVI} line. The {\fuse} telescope had relatively low throughput ($\sim$2\% end-to-end throughput), but a high enough effective area (27 cm$\rm ^{2}$) to be sensitive to the detections described in this paper. For the detections described here, the dominant noise source was the {\fuse} detector background. The sensitivity of {\fuse} to the {\OVI} line is shown in Figure \ref{fig:sensfuture}, the black dash-dotted line, with the detection limits from NGC 4631 and NGC 891 shown. 

We compare this performance with the performance of three proposed UV missions: {\luvoira}, {\luvoirb} \citep{2019arXiv191206219T}, and {\cetus} \citep{2019arXiv190910437H}. {\luvoir} (grey lines), despite its large aperture, has extremely small resolution elements from the micro-shutter array (0.054{\arcsec} $\times$ 0.108{\arcsec}; spectral $\times$ spatial). This limits the ultimate amount of light collected in any one resolution element, which is challenging for very nearby targets and yields low sensitivity, even for very large apertures. Co-adding many micro-shutter elements together provides better sensitivity (dotted grey line shows 10 resolution elements binned together). {\cetus} (red line) performs very similarly to {\fuse}, per a 2{\arcsec} $\times$ 4{\arcsec} (spectral $\times$ spatial) area on the slit. We put details of the sensitivity calculation in \autoref{app:sensitivity_calc_description}.

We contrast these mission concepts with {\aspera}, a recently funded NASA Pioneer's SmallSat (PI: Dr. Carlos Vargas). {\aspera} was explicitly designed to observe {\OVI} from nearby edge-on galaxies, and is able to reach a higher sensitivity than {\fuse}, {\luvoir}, or {\cetus} per resolution element for exposures greater than $\sim$30 hours, despite its small size and low cost. {\aspera}'s design is directly inspired by {\fuse} and uses \textit{four} off-axis parabola primary mirrors and Rowland circle spectrographs to achieve high throughput and moderate resolution while covering a large spatial area. {\aspera} is designed with a resolution element of 30{\arcsec} $\times$ 60{\arcsec} (spectral $\times$ spatial) area on the slit. {\aspera} can simultaneously observe a total area of 2{\arcmin} $\times$ 45{\arcmin} region, split into four 30{\arcsec} $\times$ 45{\arcmin} size slits. {\aspera}'s spectral resolution is $\sim$2000 at 1035 {\AA}.

\subsection{Conclusions}
In this paper, we analyze the archival {\fuse} spectra from the halo of two nearby star-forming, gas-rich, edge-on galaxies, NGC 4631 and NGC 891. We re-processed the raw {\fuse} data using the latest version of the {\fuse} data reduction pipeline. Among the five {\fuse} pointings on NGC 4631, we have found three new {\OVI} emission detections and confirmed the two previously known detections. Among the three {\fuse} pointings on NGC 891, we have found one {\OVI} emission detection, contrary to the previously reported non-detections on all three pointings. The intensity and \replaced{line-of-sight profile}{spatial distribution} of the NGC 4631 {\OVI} detections \deleted{suggest that the warm-hot halo gas could originate from different sources/mechanisms, and} cannot be explained with a simple diffuse halo model surrounding the galaxy. Our results imply that a discrete filamentary structure exists in the warm-hot phase above the central disk of NGC 4631, within a \replaced{more}{lower intensity} uniformly diffuse, regularly rotating halo. \deleted{This filament may be an example of warm-hot gas accretion onto the galaxy.} The warm-hot gas detections closest to the disk's periphery of NGC 4631 most likely represent a diffuse component that originates from the disk. A weak {\OVI} detection is also observed from one field in NGC 891. 

\deleted{It is clear that a future UV space mission with a wide field-of-view and high sensitivity is needed to determine the prevalence of circumgalactic warm-hot gas and its large-scale spatial distribution in nearby galaxies. We investigate the possibility of detecting such faint and diffuse warm-hot phase gas with three future UV missions: one NASA-funded mission in the early stage of development ({\aspera}), and two other missions in the concept development phase ({\luvoir} and {\cetus}). Among these missions, {\aspera} is specifically designed to observe {\OVI} emission, and it is expected to capture the low surface brightness {\OVI} emission with low spatial resolution but a wide field of view. Though less sensitive than {\aspera}, {\cetus} is also capable of observing {\OVI} with higher spatial resolution. {\luvoir} can observe the smallest angular scale features, but with lower sensitivity due to the small sampling size. }

Observations of {\OVI} provide a window into an important CGM phase for galaxies. \deleted{The naive assumption that a larger telescope aperture will be necessary to conduct this study or even optimized for this study is not correct. }A well designed small telescope can provide better sensitivity in a shorter time than a very large telescope which is not optimized around this type of wide field, diffuse signal. In these cases, a specifically designed mission such as {\aspera}, funded via the NASA Pioneers Program, can answer extremely important science questions at a fraction of the cost of a Flagship mission.

\section*{Acknowledgement}
We thank the anonymous reviewer for their comments on this work. 

\bibliography{ms}

\clearpage
\appendix
\renewcommand\thetable{\thesection.\arabic{table}}
\renewcommand\thefigure{\thesection.\arabic{figure}}
\setcounter{table}{0}
\setcounter{figure}{0}

\begin{figure}
\includegraphics[width=\linewidth]{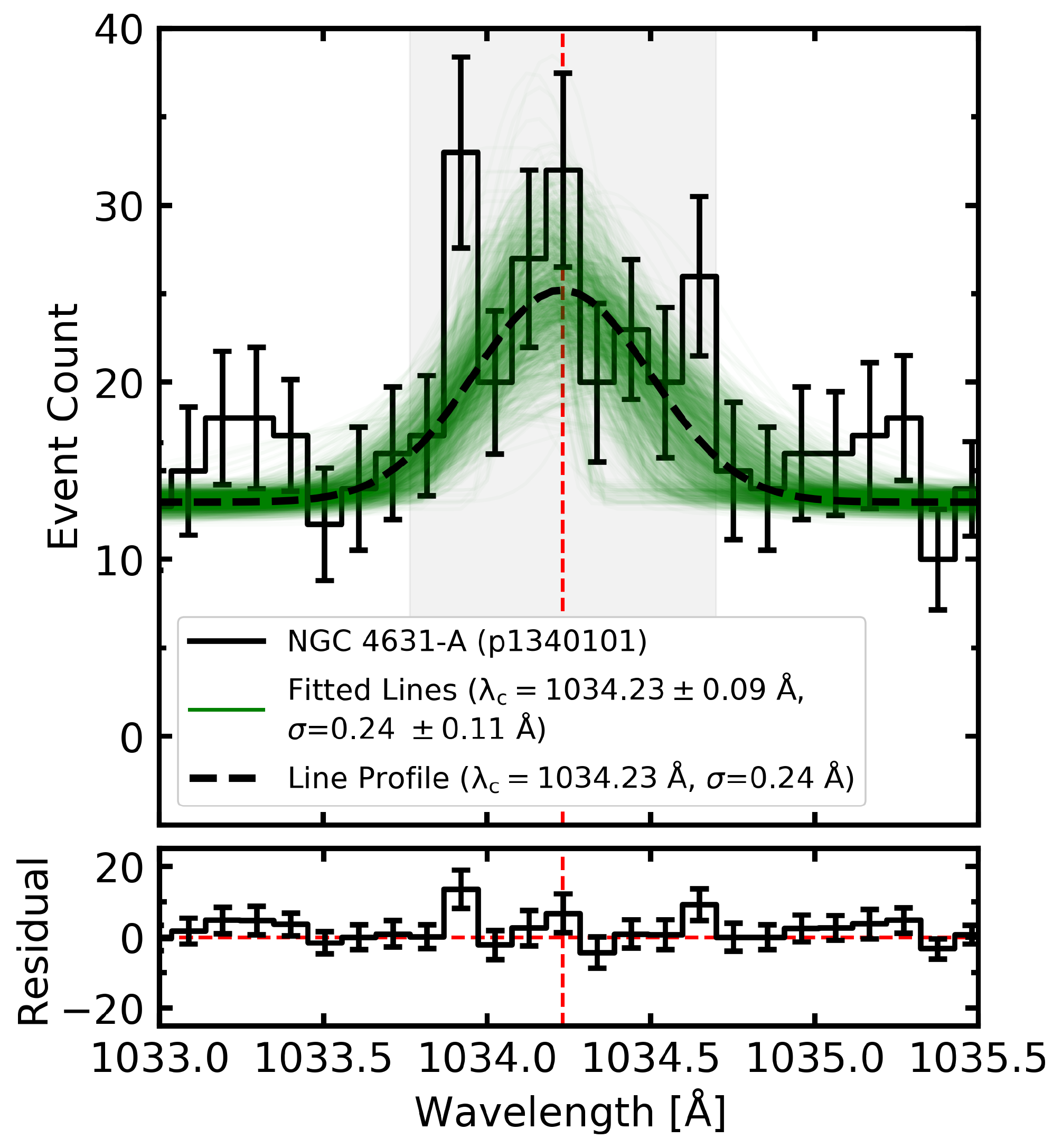}
\caption{(Top) The result of Monte-Carlo line-fitting to the {\OVI} emission in NGC 4631-A spectrum. The spectrum is shown in black with 1$\sigma$ noise range. The 1000 overlaid green lines are line profiles fitted to the 1000 random Poisson-noise-added NGC 4631-A spectra. The black dashed line is the profile with mean line parameter values. Grey region indicates line emission extraction window. (Bottom) The residual spectrum between the NGC 4631-A spectrum and the line profile with the average parameter values.
}\label{fig:line_fitting}
\end{figure}

\added{
\section{Line Fitting}\label{app:line_fitting}
It is important to carefully estimate the uncertainty of fitted parameter, when the fitting is attempted on the low signal to noise ratio data. To estimate the error of the fitted parameter, we created 1000 Poisson random noise added spectra, fit the line function to each spectrum, and take a mean and standard variation of parameters from the 1000 fittings. Although \texttt{CalFUSE} returns flux and error, generating random noise using Gaussian statistics is not advised because Poisson statistics cannot be approximated to Gaussian statistics when the signal is low. Therefore, we generated the random noise added spectra by using the \texttt{COUNTS} spectrum (Raw counts in extraction window, Table 4.8 of {\fuse} data handbook\footnote{https://archive.stsci.edu/fuse/dh.html}). The Poisson noise added spectrum is further scaled by the ratio between \texttt{COUNTS} and \texttt{WEIGHT} spectrum and used as an input for the fitting. \texttt{NOISE} spectrum is determined by the square root of \texttt{COUNTS} spectrum scaled by the ratio between \texttt{COUNTS} and \texttt{WEIGHT} spectrum.
As described in \autoref{subsec:analysis}, the line profile function determined by four parameters (Amplitude and the FWHM of the {\OVI} emission model, a center of the convolved function, and a constant) is fitted to each random-noise added spectrum. The fitting range is limited to 1030 {\AA} to 1038 {\AA}. For the line fitting, we assumed a flat background, represented by a constant as one of the four parameters.  
In \autoref{fig:line_fitting}, we put the distribution of 1000 line profiles fitted to the 1000 Poisson random noise added NGC 4631-A spectra. A line profile with the mean parameter values is plotted in black dashed line. The residual spectrum is shown in the bottom panel. 
}

\section{Sensitivity Calculation}\label{app:sensitivity_calc_description}

\begin{deluxetable*}{
>{\centering}p{0.255\textwidth}
>{\centering}p{0.105\textwidth}
>{\centering}p{0.105\textwidth}
>{\centering}p{0.105\textwidth}
>{\centering}p{0.105\textwidth}
c
}
\tabletypesize{\footnotesize}
\tablewidth{0pt}
 \tablecaption{SNR Calculation Breakdown per sampling element\tablenotemark{a}  at 1035 {\AA}  (Exposure time: 388800 sec)  \label{tab:snrbreakdown}}
 \tablehead{ 
\colhead{\multirow{2}{*}{Parameters} }
 &  \colhead{\specialcell[t]{{\luvoira} \\ (LUMOS, 155L)}} 
 & \colhead{\specialcell[t]{{\cetus} \\ (LUV G117)}} 
 & \colhead{\specialcell[t]{{\fuse} \\ (LWRS, LiF1-A)}} 
 & \colhead{\multirow{2}{*}{{\aspera}} }
& \colhead{\multirow{2}{*}{Unit} }
}
 \startdata  
\multirow{2}{*}{O VI 1032 line intensity} &       21800 &      1550 &      1270 &       827 &   counts s$^{-1}$ cm$^{-2}$ sr$^{-1}$ \\
$\;$                                      &    9.83E-18 &  7.01E-19 &  5.72E-19 &  3.73E-19 &  erg s$^{-1}$ cm$^{-2}$ arcsec$^{-2}$ \\
Effective area                            &     7.1E+04 &   1.1E+03 &        27 &       2.1 &                                cm$^2$ \\
Full slit length                          &          91 &       120 &        30 &      2700 $\times$ 4\tablenotemark{b} &                                arcsec \\
Spatial sampling        &       0.110 &         4 &        30 &        60 &                                arcsec \\
Slit width           &       0.055 &         2 &        30 &        30 &                                arcsec \\
Total background rate at MCP              &        1.05 &      0.30 &      0.96 &      0.30 &             counts s$^{-1}$ cm$^{-2}$ \\
Extraction aperture size (Spatial)        &      0.0159 &    0.0508 &    0.0800 &    0.0047 &                                    cm \\
Extraction aperture size (Spectral)       &      0.0292 &    0.0844 &    0.0960 &    0.0087 &                                    cm \\
Total signal count                        &          82 &       125 &       281 &        29 &                                counts \\
Noise (MCP Background)                    &          14 &        22 &        54 &         2 &                                counts \\
Noise (Total RSS)                         &          16 &        25 &        56 &         6 &                                counts \\
\hline
Total Signal to Noise Ratio               &         5.0 &       5.0 &       5.0 &       5.0 &                                       \\
Grasp\tablenotemark{c}                    &        98.4 &      73.3 &       6.7 &     192.6 &                 cm$^{2}$ arcmin$^{2}$  \\
\enddata
\tablenotetext{a}{Spatial sampling $\times$ slit width}
\tablenotetext{b}{Four 45{\arcmin} $\times$ 30{\arcsec} slits}
\tablenotetext{c}{Effective area $\times$ full slit length $\times$ slit width}
\end{deluxetable*}

We calculate diffuse {\OVI} 1032 {\AA} emission line sensitivity of the past ({\fuse}) and future ({\luvoira}, {\luvoirb}, {\cetus}, and {\aspera}) UV missions. The detector of all four missions is micro-channel plate (MCP). The lines are assumed to be slightly redshifted to 1035 {\AA}. We have considered only two sources of noise for the calculation: target shot-noise and intrinsic MCP background noise, because those two are the dominant sources of noise at the wavelength around 1030 {\AA}. Other sources of noise could be earthshine (not applicable for {\luvoir} or {\cetus}), Zodiacal light, or instrument-specific scattered/stray light, but their contribution to the noise would be negligible at this wavelength range. Earthshine (assuming the Earth limb angle $>$ 15$^{\circ}$) and Zodiacal light are considerably dimmer than 10$^{-21}$ {\fluxerg} (COS Instrument Handbook\footnote{https://hst-docs.stsci.edu/cosihb}). Scattered/stray light is assumed to be well-managed with quality optics, proper opto-mechanical arrangements and baffling. With the two kinds of noise sources, the sensitivity is calculated as a function of exposure time. The result is presented in \autoref{fig:sensfuture}. Representative SNR breakdown is shown in \autoref{tab:snrbreakdown} for the case of {\luvoira}, {\cetus}, {\fuse}, and {\aspera} at 1035 {\AA} with the exposure time of 388,000 sec (4.5 days, the nominal total exposure time per exposure of {\aspera} mission). The grasp (effective area $\times$ field-of-view) of the four missions are also shown in \autoref{tab:snrbreakdown}.

Below we describe the justification of adopted parameters. The resulting sensitivity curves are the best estimates based on the available information and will be different if there are changes to the parameter values. As guidance for the impact of changes to the sensitivity, we have checked the sensitivity curve of the missions with two additional cases, (1) 20\% larger effective area with 20\% smaller background noise and (2) 20\% smaller effective area with 20\% larger background noise with respect to the baseline values. The sensitivity will be increased by 0.09-0.13 dex for case (1), and decreased by 0.11-0.13 dex for case (2).

\subsection{LUVOIR}
The instrument parameters of {\luvoira} and B are estimated from the {\luvoir} mission concept study final report \citep{2019arXiv191206219T}. LUMOS-G155L grating is chosen because the grating is expected to have the highest sensitivity to the {\OVI} emission while provide a reasonable diffuse-source spectral resolution (R$\sim$2500), assuming a fully-illuminated micro shutter (0.110{\arcsec}$\times$0.055{\arcsec}). LUMOS can open 840 micro shutters simultaneously within the 2$\arcmin$ $\times$ 2$\arcmin$ micro shutter array field of view. The effective area of G155L grating at 1035 {\AA} is estimated by assuming the spectral shape of the effective area curve from Figure 8 of \citet{2017SPIE10397E..13F}, scaled by the peak effective area value of G155L grating from Table 8-9 of the {\luvoir} final report \citep{2019arXiv191206219T} for both {\luvoira} and {\luvoirb} concepts. MCP background rate of 1.05 counts s$^{-1}$ cm$^{-2}$ is taken from \citet{2017SPIE10397E..13F}. The extraction aperture size (293 $\mu$m $\times$ 159 $\mu$m) corresponds to the size of fully-illuminated un-vignetted open area of a single microlens shutter size (corresponds to 0.054$\arcsec$ $\times$ 0.108$\arcsec$ are on sky, physically 79 $\mu$m $\times$ 159 $\mu$m) with 1 {\AA} width on the detector. This assumes unit magnification of LUMOS FUV G155L optics.

\subsection{{\cetus}}
{\cetus} instrument parameter values are adopted from {\cetus} final report \citep{2019arXiv190910437H}.
Point/Slit spectrograph (PSS) with LUV G117 grating mode is chosen because this is the only available {\cetus} configuration to observe local {\OVI} 1032 {\AA} emission line. The effective area of {\cetus} G117 configuration at 1035 {\AA} is estimated from Figure 1-3 of \citet{2019arXiv190910437H}. MCP background rate is assumed as 0.3 counts s$^{-1}$ cm$^{-2}$. The MCP background rate is estimated by the summation of the atomic layer deposited (ALD) MCP background rate measured at the ground (0.05 counts s$^{-1}$ cm$^{-2}$) and the expected difference between the flight and ground background rate with ALD MCP with satellite mass of $\sim$2.6 metric ton (0.25 counts s$^{-1}$ cm$^{-2}$, \citet{Siegmund2020}). The extraction aperture size on the detector (844 $\mu$m $\times$ 508 $\mu$m) corresponds to the size of {\OVI} emission with 1 {\AA} width, fully-illuminated 2{\arcsec} $\times$ 4{\arcsec} (spectral $\times$ spatial) area on the slit. 1:2 ratio of the sampling element size is chosen arbitrarily to be aligned with the sampling shape of {\luvoir} micro shutter. The length of {\cetus} PSS LUV slit is 2{\arcmin}, so 30 such sampling elements can be observed simultaneously.

\subsection{{\fuse}}
Although {\fuse} is already retired, an estimation with {\fuse} is also shown as a reference. Here we took the instrument parameters as per the NGC 4631-A observation (Effective area=26.9 cm$^2$, Total background rate=0.96 counts s$^{-1}$ cm$^{-2}$) and the {\fuse} instrument handbook\footnote{https://archive.stsci.edu/fuse/ih.html}. The field of view of LWRS is 30{\arcsec} $\times$ 30{\arcsec}. The extraction aperture size corresponds to the footprint of fully illuminated LWRS on the detector at around 1035 {\AA} (Height: 800 $\rm \mu$m; identified from detector image, width: 960 $\rm \mu$m; corresponds to 1.08 {\AA}).

\subsection{{\aspera}}
{\aspera} instrument parameters are adopted from {\aspera} proposal that was submitted to 2020 NASA Pioneers Announcement of Opportunity (Vargas et al. Private Comm.). Effective area of 2.1 cm$^2$ at 1035 {\AA} and total background rate of 0.3 counts s$^{-1}$ cm$^{-2}$ are used for the sensitivity calculation. The extraction aperture size (47 $\mu$m $\times$ 87 $\mu$m) on the detector corresponds to the size of 1 {\AA} width {\OVI} emission with fully-illuminated 30{\arcsec} $\times$ 60{\arcsec} (spectral $\times$ spatial) area on the slit. By employing the step-and-stare concept observation, {\aspera} can effectively map the spatial distribution of {\OVI} emission in galaxy halos.

\end{document}